\documentclass[aps,a4paper,twocolumn]{revtex4}

\usepackage{epsfig}
\usepackage{graphicx}
\usepackage{amsmath,amssymb,color}
\usepackage[english]{babel}
\usepackage{graphicx}

\parskip=\medskipamount

\def\beq{\begin{equation}}
\def\eeq{\end{equation}}
\def\beqn{\begin{eqnarray}}
\def\eeqn{\end{eqnarray}}

\def\dim{{\mathrm{dim}}}

\newcommand{\eq}[1]{(\ref{#1})}
\newcommand{\fig}[1]{Fig.\ref{#1}}

\def\disp{\displaystyle}

\def\la{\left<}
\def\ra{\right>}

\begin{document}

\title{Critical Behavior in Topological  Ensembles}

\author{K. Bulycheva$^{1,5}$, A. Gorsky$^{1,2}$, and S. Nechaev$^{3,4}$}

\affiliation{$^1$Institute of Information Transition Problems, B.Karetnyi 19, Moscow, Russia \\
$^2$Moscow Institute of Physics and Technology, Dolgoprudny 141700, Russia \\
$^3$Universit\'e Paris-Sud/CNRS, LPTMS, UMR8626, B\^at. 100, 91405 Orsay, France \\
$^4$P.N.Lebedev Physical Institute, RAS, 119991 Moscow, Russia \\
$^5$Department of Physics, Princeton University, USA}


\begin{abstract}

We consider the relation between three physical problems: 2D directed lattice random walks,
ensembles of $T_{n,n+1}$ torus knots, and instanton ensembles in 5D SQED with one compact dimension
in $\Omega$ background and with 5D Chern-Simons term at the level one. All these ensembles exhibit
the critical behavior typical for the "area+length+corners" statistics of grand ensembles of 2D
directed paths. Using the combinatorial description, we obtain an explicit expression of the
generating function for $q$-Narayana numbers which amounts to the new critical behavior in the
ensemble of $T_{n,n+1}$ torus knots and in the ensemble of instantons in 5D SQED. Depending on the
number of the nontrivial fugacities, we get either the critical point, or cascade of critical lines
and critical surfaces. In the 5D gauge theory the phase transition is of the 3rd order, while in
the ensemble of paths and ensemble of knots it is typically of the 1st order. We also discuss the
relation with the integrable models.

\end{abstract}


\maketitle

\section{Introduction}

Challenging questions appear often at edges of traditional fields. As an example, the new branch of
mathematical physics, the "statistical topology" emerged recently by absorbing ideas from the
statistical physics, theory of integrable systems, and algebraic topology (see, \cite{nechaev} for
review). The scope of the statistical topology includes, on the one hand, mathematical problems
involved in the construction of topological invariants of knots and links based on solvable models
and, on the other hand, the physical and statistical problems related to summation over knot
ensembles. In this work, we dwell predominantly to problems of the latter kind, demonstrating the
emergence of a critical behavior in ensemble of $T_{n,n+1}$ torus knots. This critical behavior is
formulated in terms of knot invariants.

\begin{figure}[ht]
\centerline{\includegraphics[width=8cm]{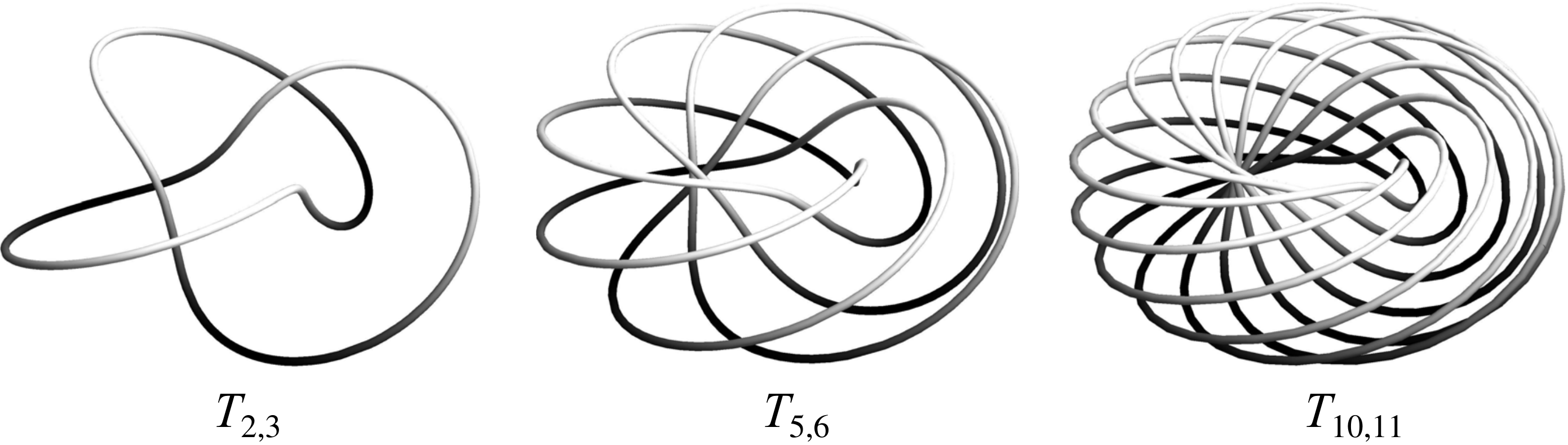}}
\caption{Few samples of torus knots from the series $T_{n,n+1}$: $T_{2,3},\; T_{5,6},\; T_{10,11}$.}
\label{fig:01}
\end{figure}

Torus knots $T_{m,n}$ seem to be among the simplest objects in the knot theory. It is difficult to
overestimate their role in different branches of mathematical physics. The topology of a torus knot
is uniquely determined by the pair $(m,n)$, which fixes windings along two torus periods. In the
\fig{fig:01} few particular examples of torus knots, $T_{2,3},\; T_{5,6},\; T_{10,11}$ from the
series $T_{n,n+1}$ are depicted. The closed curves wrap around the torus which is not shown.

Various explicit expressions for knot invariants are known, ranging from classical Jones--Kauffman
polynomials \cite{kauffman,russo}, to recent superpolynomials of torus knots \cite{gn}. New
approaches to the construction of torus knot invariants for particular knots, based on the
application of topological string theories and deformed matrix models, have been formulated
relatively recently in \cite{as,vafa,marino}. Much less is known about properties of \emph{knot
ensembles}, where the particular topology of a knot diagram is considered as a topologically
"quenched" variable similar to the quenched disorder in statistical physics. The weighted summation
over different torus knot types (i.e. different pairs $(m,n)$) means the consideration of the grand
canonical ensemble (i.e. of the generating function) of torus knots.

\begin{figure}[ht]
\centerline{\includegraphics[width=8cm]{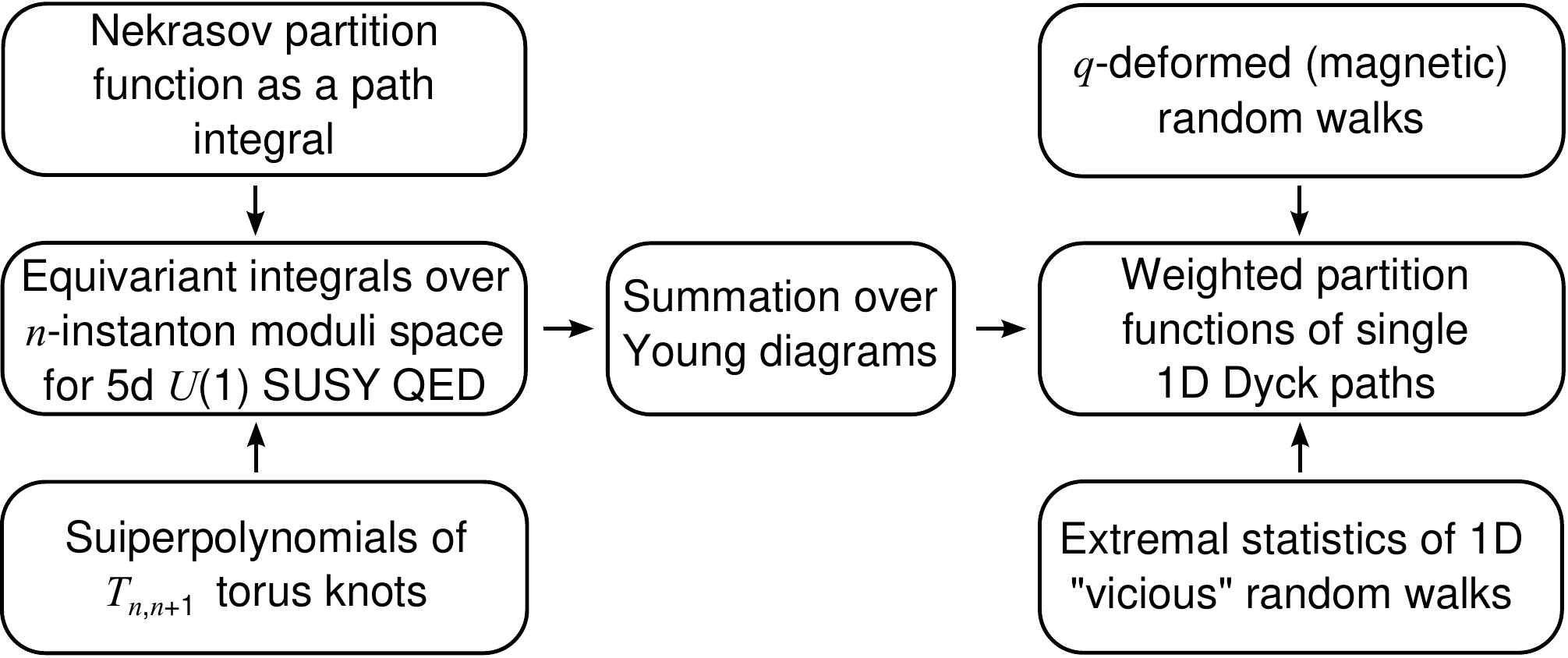}}
\caption{Schematic connection between different systems discussed in at length of the paper.}
\label{fig:02}
\end{figure}

In this work we uncover the relation between three physical problems: i) two-dimensional directed
(i.e. (1+1)-dimensional) lattice random walks with fixed area under the curve, ii) ensemble of
$T_{n,n+1}$ torus knots, and iii) a five-dimensional SQED with the Chern--Simons term at the level
one. The reason for random walks to appear in ii) and iii) can be intuitively explained as follows.
The main tool for the evaluation of the torus knot superpolynomials \cite{gn} and of Nekrasov
partition function in the SUSY gauge theory \cite{nekrasov}, is the "equivariant localization"
approach, which reduces the integral over the particular moduli space to the summation over the
Young tableau. The last problem can be reformulated as a weighted sums over directed paths on a
square lattice. The schematic relations between the problems considered in the paper is shown in
the flowchart in the \fig{fig:02}.

Our main goal is the investigation of the critical behavior in these topological ensembles. From
the "random walk side" the critical behavior occurs in the space of fugacities and we shall focus
at the "area+length+corners" statistics of paths. We derive an explicit expression for the free
energy in the random walks problem for the "area+length+corners" statistics, which provides the
very nontrivial phase structure in the three-dimensional space of fugacities. In the generic case
we find a cascade of phase transitions and in the degenerate cases we reproduce the previously
known results.

The relation between the random walks and the torus knot superpolynomials can be traced from the
mathematical literature, however the relation with the particular observable in 5D SQED instanton
ensemble is new. Collecting different points of view, and the results of \cite{gm}, we interpret
the generating function of ensemble of $T_{n,n+1}$ knots as the weighted sum of instanton
contributions to the particular observable in 5D SQED and analyze the structure of corresponding
generating function. The interpretation of the random walk fugacities in terms of the generating
parameters in the torus knot ensemble is quite straightforward, moreover, these fugacities are
identified in the 5D SQED as well: the corresponding parameters turn out to be the gauge coupling,
the mass of the hypermultiplet, and the parameters of the $\Omega$-deformation.

Given an explicit expression for the free energy in the random walks problem for the
"area+length+corners" statistics, we use it to analyze the ensembles of knots and instantons. We
show that at the "gauge theory side" the particular third derivative of the instanton partition
function with respect to the masses exhibits an unexpected critical behavior at some critical line
in the space of parameters. We provide a physical interpretation of critical behaviors in all three
theories considered here. Having an exact expression for the $q$-Narayana numbers (the generating
function of the area- and corner-weighted (1+1)D Brownian excursion), we describe explicitly the
phase portrait of different ensembles. We show that there is a 3rd order phase transition in the
instanton ensemble, corresponding to the 1st order phase transition in the ensembles of random
paths and torus knots.

The paper is organized as follows. In the Section 2 we describe the sum over the paths with
different statistics and focus at the "area+length+corners" ensembles. In the Section 3 we discuss
the representation of the partition function of (1+1)D Brownian excursions (Dyck paths) for the
superpolynomials of $T_{n,n+1}$ torus knots. The Section 4 is devoted to the identification of the
sum over the paths as the instanton contribution to the particular observable in the 5D SQED. In
the Section 5 we consider the physical interpretation of the critical behavior in the space of
fugacities in all models discussed at length of the paper. The relation with the Toda-like
integrable system is mentioned in Section 6. Our findings and the open questions are summarized in
the Conclusion. In the Appendix A, using the combinatorial description, we derive the new explicit
expression for the generating function of $(q,a,s)$-Narayana polynomials (the generating function
of area- and corner-weighted Dyck paths). In the Appendix B we remind for completeness the relation
of the critical behavior with the "hydrodynamic description" of the edge singularities in GUE
matrix ensembles.

\section{Critical behavior of area- and corner-weighted Dyck paths}

\begin{figure}[ht]
\centerline{\includegraphics[width=6cm]{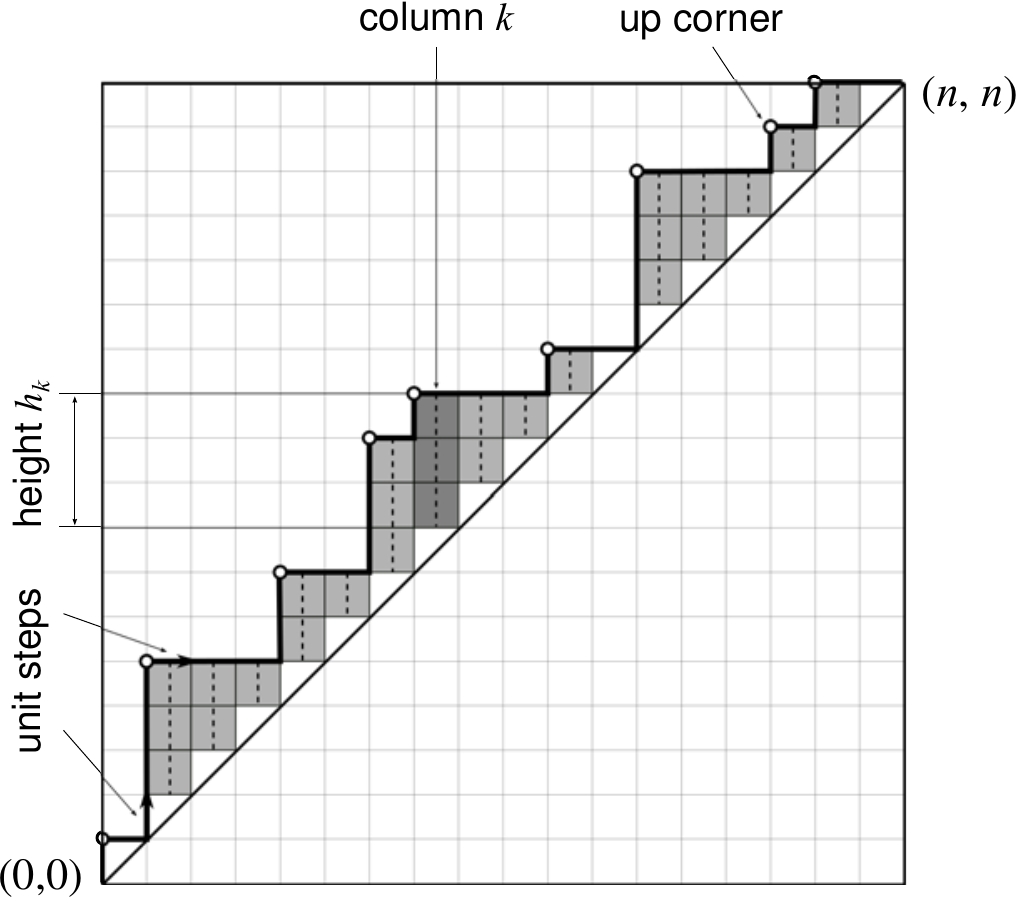}}
\caption{Sample of a Dyck path (Brownian excursion above the diagonal of the square) with a fixed
area between the path and diagonal counted in full plaquettes (grey boxes) and a fixed number of
up-corners (local "peaks" shown by open dots). The partition function of such paths is given by a
generalization of $q$-Narayana polynomials.}
\label{fig:03}
\end{figure}

By definition, the Dyck path of length $2n$ on the square lattice starts at the origin $(0,0)$,
ends at point $(n,n)$ and consists of the union of sequential elementary "$\uparrow$" and "$\to$"
steps, such that the path always stays above the diagonal of the square -- see the \fig{fig:03}.
The number of all Dyck paths of length $2n$ is given by the Catalan number, $C_n=\frac{\disp
1}{\disp n+1}\left(\begin{matrix} 2n \\ n\end{matrix}\right)$. We also denote Dyck paths as
"Brownian excursions" (BE), having in mind an image of a charged particle on a square lattice in an
external transversal magnetic field (after applied Wick rotation), where the motion of a particle
is subject to two restrictions: it moves only up and right and never intersects the diagonal.
Calculating the action for such a particle, we see that $q=\exp(i \rm external\, magnetic\, field)$
is the fugacity of the area, $A$, under the Dyck path, and the $s=\exp(\rm mass)$ is the fugacity
of the path length, $n$. The information about statistics of area-weighted Dyck paths can be easily
extracted from the generating function, which is the sum over all path lengths. This model can be
referred to as the "chiral Hofstadter system", considered in \cite{ouvry}.

\begin{figure}[ht]
\centerline{\includegraphics[width=8cm]{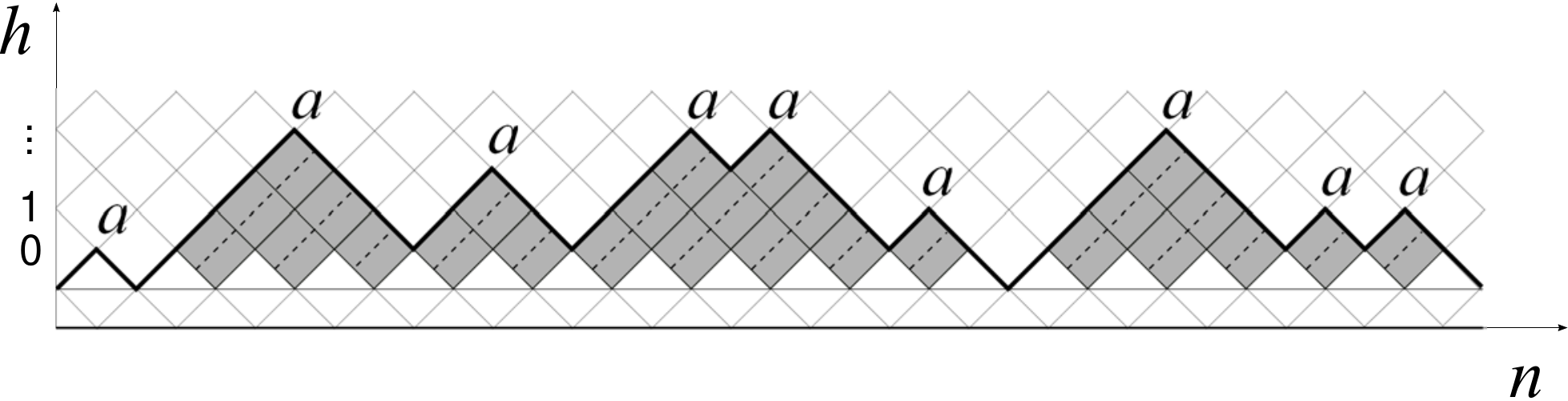}}
\caption{(a) The $2n$-step Dyck path (the same figure as in the \fig{fig:03}) turned by $\pi/2$,
the heights are measured along the dotted lines and the $\wedge$-corners have fugacity $a$.}
\label{fig:04}
\end{figure}

To proceed, turn the lattice by $\pi/4$ and write the recursion relation for the partition function
$Z_k(x;q)$ on a half-line, $x\ge 0$, where $x$ is the height of the path at the step $k$. The area,
$A$, below the path is counted as a sum of filled plaquettes (i.e. "heights") between the path and
the $x=0$-axis, as shown in the \fig{fig:04}a. Each plaquette has the weight $q$.

The partition function of area-weighted Dyck paths, $Z_k(x;q)$, satisfies the relation
\beq
\left\{\begin{array}{l} Z_{k+1}(x;q) = q^{x-1} Z_k(x+1;q) + Z_k(x-1;q) \medskip  \\
Z_k(0;q)=0 \medskip \\
Z_{k=0}(x;q)=\delta_{x,1}
\end{array}\right.
\label{eq:01}
\eeq
Equivalently, \eq{eq:01} can be written in a matrix form. Define ${\bf
Z}_n(q)=\big(Z_n(1;q),Z_n(2;q),Z_n(3;q),...\big)^{\intercal}$. Then
\beq
{\bf Z}_n(q) = T^n(q) {\bf Z}_0
\eeq
where
\beq
T(q)=\left(\begin{array}{ccccc}
0 & 1 & 0 & 0 & \ldots \\
1 & 0 & q & 0 & \ldots \\
0 & 1 & 0 & q^2 &  \\
0 & 0 & 1 & 0 &  \\
\vdots & \vdots &  &  & \ddots \end{array}\right); \quad {\bf Z}_0=\left(\begin{array}{c} 1 \\0 \\
0 \\ \vdots \end{array}\right)
\label{eq:02}
\eeq
We are interested in the value $Z_n(1;q)$ since at the very last step the trajectory returns to the
initial point. Evaluating powers of the matrix $T$, we can straightforwardly check that the values
of $Z_N(1;q)$ are given by Carlitz--Riordan $q$-Catalan numbers, namely,
\beq
Z_n(1;q) = \left\{\begin{array}{ll} C_{n/2}(q) & \mbox{for $n=2k$ ($k=1,2,...$)}
\medskip \\ 0 & \mbox{for $n=2k-1$ ($k=1,2,...$)}
\end{array} \right.
\label{eq:03}
\eeq
Recall that $C_n(q)$ satisfy the recursion
\beq
C_n(q) = \sum_{k=0}^{n-1}q^k C_k(q) C_{n-k-1}(q)
\label{eq:04}
\eeq
which is the $q$-extension of the standard recursion for Catalan numbers. The generating function
$$
F(s,q)=\sum_{n=0}^{\infty}s^n C_n(q)
$$
obeys the functional relation
\beq
F(s,q) = 1 + s F(s,q) F(sq,q)
\label{eq:05}
\eeq
It is known that the solution of \eq{eq:05} can be written as a continued fraction expansion,
\beq
F(s,q) =\frac{1}{\disp 1-\frac{s}{\disp 1-\frac{s q}{\disp 1- \frac{s q^2}{1-...}}}}
=\frac{A_q(s)}{A_q(s/q)}
\label{eq:06}
\eeq
where $A_q(s)$ is the $q$-Airy function,
\beq
A_q(s)=\sum_{k=0}^{\infty}\frac{q^{k^2}(-s)^k}{(q;q)_k}; \quad (t;q)_k=\prod_{k=0}^{k-1} (1-t q^k)
\label{eq:07}
\eeq

Let us describe how the critical behavior emerges at the Brownian excursion side. In the works
\cite{prel0,rich1, rich2} it has been shown that in the double scaling limit $q\to 1^-$ and $s\to
\frac{1}{4}^{-}$ the function $C(s,q)$ has the following asymptotic form
\beq
C(z) \sim C_{\rm reg}+(1-q)^{1/3} \frac{d}{dz}\ln {\rm Ai}(4z); \quad
z=\frac{\frac{1}{4}-s}{(1-q)^{2/3}},
\label{eq:asymp1}
\eeq
where $C_{\rm reg}$ is the regular part at $\big(q\to 1^-,\, s\to \frac{1}{4}^{-}\big)$ and $\disp
{\rm Ai}(z)=\frac{1}{\pi} \int_{0}^{\infty} \cos(\xi^3/3+\xi z)\, d\xi$ is the Airy function.

The function $C(s,1)$ is the generating function for the undeformed Catalan numbers:
\beq
C(s,q=1)=\frac{1-\sqrt{1-4s}}{2s}
\eeq
The generating function $C(s,1)$ is defined for $0<s<\frac{1}{4}$, and at the point $s=\frac{1}{4}$
the first derivative of $C(s,1)$ experiences a singularity which is interpreted as the critical
behavior. The limit $q=1$, $s\to\frac{1}{4}^-$ can be read also from the asymptotic expression for
$C(s,q)$, Eq.(\ref{eq:asymp1}):
\beq
C(s,q)\big|_{q\to 1^-}\sim C_{\rm reg}-2\sqrt{1-4s}.
\label{eq:limq1}
\eeq
where $C_{\rm reg}$ is the regular part of $C(s,q)$ at $q\to 1$ and $s\to \frac{1}{4}$ and $C_{\rm
reg}=2$ at $s=\frac{1}{4}$. Note that the first non-singular term in (\ref{eq:limq1}) does not
contain $q$, so it is no matter in which order the limit in (\ref{eq:asymp1}) is taken. However to
define the double scaling behavior and derive the Airy-type asymptotic, the simultaneous scaling in
$s$ and $q$ is required.

The generating function, $F(a,s)$, for Narayana numbers, which count Dyck paths with fixed fugacity
of corners, $a$, demonstrates the behavior similar to \eq{eq:limq1}, namely the square-root
singularity,
\beq
F(a,s)= \frac{1-(1-a)s-\sqrt{(1+s-sa)^2-4s}}{2s}
\label{eq:nar}
\eeq
This behavior at $a=1$ coincides with the one of Catalans \eq{eq:limq1}. In Appendix A we have
derived the explicit expression for the generating function $F(q,a,s)$ of $q$-Narayanas with the
Airy-type asymptotic \eq{eq:asymp1} in terms of $q$-orthogonal polynomials related to
Rogers--Ramanujan continued fractions,
\beq
F(q,a,s)=\frac{A_q(s;s(1-a))}{A_q(s/q;s(1-a)/q)}
\label{eq:crit}
\eeq
where $A_q(s;s(1-a))$ is the extension of the $q$-Airy function $A_q(s)$ defined in \eq{eq:07}. The
function $A_q(s;s(1-a))$ reads
\beq
A_q(s;s(1-a)) = \sum_{k=0}^{\infty} \frac{q^{k^2}(-s)^k}{(q;q)_k\,(-s(1-a);q)_k}
\label{eq:crit2}
\eeq
One can immediately see that $A_q(s,s(1-a))\big|_{a=1} = A_q(s)$, where $A(s)$ is given by
\eq{eq:07}.

Completing this Section it is worth reminding that appearance of the singularity of type
\eq{eq:asymp1} is the manifestation of the third-order phase transition. In the seminal paper
\cite{BDJ} it has been shown that the largest eigenvalue, $\lambda_n$, of the Gaussian $n\times n$
random matrix ensemble, converges at $n\to\infty$ to $\lambda_n \to 2\sqrt {n} + n^{1/6} \chi$,
where the random variable $\chi$ has a limiting $n$-independent distribution, ${\rm Prob}(\chi\leq
x) = F_{\rm GUE}(x)$, being the so-called Tracy-Widom distribution for GUE ensemble \cite{TW}. So,
the normalized value $\Lambda_n= \lambda_n/\sqrt{n}$ at large (but finite) $n$ has an uncertainty
(i.e. the width of the distribution) of order of $n^{-1/3}$, typical for the 3rd order phase
transitions. Above and below the critical value $\Lambda_{\infty}=\lim\limits_{n\to\infty}
\Lambda_{n}=2$, the tails of the distribution $P(\Lambda)$ have different asymptotics, signifying
existence of strong (for $\Lambda<\Lambda_{\infty}$) and weak (for $\Lambda>\Lambda_{\infty}$)
couplings.

\section{Dyck paths generating functions and torus knot invariants}

In this Section we briefly explain the relation between the invariants of knots and the random
walks with three fugacities. We restrict ourselves by the torus knots, $T_{n,m}$, and consider the
superpolynomial introduced in \cite{gukov}. They depend on three generating parameters, which can
be related to the fugacities of 2D directed random walk. Remind that the superpolynomial is the
Poincar\'e polynomial of the triply graded Khovanov homologies, $H_{ijk}$, which in the vector
space can be attributed to the knot. The knot superpolynomials are the generalizations of the
HOMFLY knot polynomials and depend on three variables, corresponding to gradings \cite{gukov}
\beq
P_{n,m}(a,q,t)=\sum_{ijk}a^i q^j t^k\; \dim\, H_{ijk}
\eeq
At $t=q^{-1}$ the superpolynomial $P_{n,m}(a,q,t)$ reduces to the standard HOMFLY polynomial.
Alternatively, it can be interpreted as the generating function for the multiplicities in the
particular sector of BPS states in the SUSY gauge theories \cite{gukov}.

We will be interested in the critical behavior of the "area+length+corner" type statistics in the
ensemble of torus knots and focus on the particular series of $T_{n,n+1}$ uncolored knots
parameterized by one integer, $n$. The main object, as before, is the generating function for
superpolynomials, $Z(s,a,q,t)$, in the ensemble of $T_{n,n+1}$ knots, where the fugacity, $s$, is
conjugated to the index $n$, which defines the winding around the cycle on the solid torus (see the
\fig{fig:01}),
\beq
Z(s,a,q, t)=\sum_{n=0}^{\infty} P_{n,n+1}(a,q,t) s^n
\eeq

Hopefully there is an explicit expression for the superpolynomial of the $T_{n,n+1}$ torus knots
obtained in two different ways. The first one deals with the combinatorics of the Young diagrams
and can be related to the Brownian excurion approach \cite{appendix}, while the second approach has
been developed in \cite{gn} via the localization on the fixed points of the torus action in the
moduli space of $n$ points in $\mathbb C^2$. The second approach will be used in the next Section
to compare the generating function for superpolyniomials of $T_{n,n+1}$ uncolored torus knots with
the instanton contribution to the particular observable in 5D SQED with the Chern--Simons term.

We focus on the BE representation and provide a dictionary identifying the fugacities at BE side
with the ones at the knot side. This dictionary establishes the translation of the BE language to
the language of torus knot superpolynomials. The construction of torus knot invariants involves two
more statistics for Dyck paths, the statistics of "up-corners", where the path changes direction
from "up" ($\uparrow$) to "right" ($\to$), and the statistics of "dinv", a definition for which can
be found in \cite{appendix}. The superpolynomial for $T_{n,n+1}$ knots expressed in terms of Dyck
paths can be presented by the following partition function:
\beq
Z_n(q,t,a) = \sum_{\pi_n\in \rm{Dyck\ paths}} q^{A}\; t^{\#\,\rm dinv}\; a^{\#\,\rm corners},
\label{eq:1}
\eeq
where $q$, $a$ and $t$ are the fugacities of area, corners and dinv correspondingly. The main
object of our study is the generating function,
\beq
Z(q,t,a,s) = \sum_{n=0}^{\infty} Z_n(q,t,a) s^n
\label{eq:1a}
\eeq
where s is fugacity for the length of the path conjugated to the integer $n$, which weights knot
type (at a knot side) and Dyck path length (at a BE side).

As before, the area, dinv, and corner statistics in the ensemble of lattice paths are represented
by the corresponding fugacities in the grand ensemble. To use the known critical behavior for the
generating function of superpolynomials in the ensemble of torus knots, we have to consider the
reduction of paths statistics to the "length+area+corners" one, switching off the "dinv" fugacity
(i.e. setting $t=1$ in \eq{eq:1}).

Due to $q\leftrightarrow t$ duality, one can equivalently consider the "length+area+corners+bounce"
statistics, switching off the "bounce" fugacity. This duality can be seen, for instance, in the
lowest row in the expansion in the $a$ variable of the superpolynomial:
\beq
P_{n,n+1}(a,q, t)= \sum_k a^k P^{k}_{n,n+1}(q,t).
\eeq
It turns out that the corresponding term at $a=0$ in the expansion coincides with the
$(q,t)$-Catalan number, namely, $P^0_{n,n+1}(q, t)= C_n(q,t)$, when the duality is well-known.

In the next Section we explain the relation between the generating function for the torus knot
superpolynomials and the some observables in the 5D SQED upon the particular identification of
parameters. In the Section 6 we shall use the generating function for $q$-Narayanas to analyze the
critical behavior in the torus knots ensembles classified by the HOMFLY knot invariants.

\section{Towards the Critical behavior in 5D SQED}

Consider now the Nekrasov-like partition function \cite{nekrasov} in the Abelian 5D SUSY gauge
theory with the  massless hypermultiplet in fundamental representation and massive multiplet in the
antifundamental representation in the $\Omega$--background. The coefficient in front of the
Chern--Simons term is fixed, $k=1$, the coupling constant in 5D theory is dimensionful, and the
fifth coordinate is compact. The Nekrasov partition function is trivial in this theory, however we
shall be interested in the vacuum matrix element $\la O \ra$ of the particular operator $O$. Its
evaluation involves the weighted sum of the integrals over the instanton moduli space, where the
parameters of the $\Omega$--background, $\epsilon_1, \epsilon_2$, serve as the equivariant
parameters of two torus actions for the integration over the moduli space, $M_n$, of $n$ point-like
instantons, all located at the origin. The instanton number is weighted with the counting parameter
$Q=e^{2\pi i \tau}$, where $\tau = 4\pi i \beta g^{-2}$ and $\beta$ is the radius of the compact
fifth dimension.

The desired operator $O$ can be identified as follows \footnote{We are grateful to N. Nekrasov for
important discussion on this point and crucial suggestion}. First, we have to recognize the deformed
Catalan numbers, $C_n(q,t)$, with $\la O \ra$ in the $n$--instanton sector. To this aim we use the
following important result \cite{haiman}
\beq
\chi^{T}\left({\rm Hilb}^n(\mathbb C^2,0), V \otimes \Lambda^n V\right) = C_n(q,t),
\eeq
which interprets the $(q,t)$-deformed Catalans as equivariant integrals over the moduli space of
$n$--Abelian instantons valued in the $n$th power of the $n$--dimensional tautological bundle, $V$.

Looking at the representation of the $(q,t)$--Catalans in terms of the paths on the Young tableau,
the desired operator, up to the normalization, can be written in the $n$--instanton sector as
\beq
\la O \ra_n \propto  \la \tilde{Q}{Q}\left({\rm Tr} e^{\Phi}\right)\ra _n,
\eeq
where $\tilde{Q},Q$ is the hypermultiplet, ${\rm Tr}$ substitutes the integral over the $\mathbb
C^2$ in the $\Omega$--background, and $\Phi$ is the "long scalar" in the $\Omega$--background
\cite{nekrasov}. Hence, the composite operator under consideration, is the product of local and
nonlocal 4-observables.

It is more convenient to use the equivalent, however a bit more symmetric formulation of the
desired observable in 5D SQED \cite{gm}. Consider the 5D QED with two flavors in the fundamental
representation with masses $m$, $M$, and the flavor in the antifundamental representation with a
mass $m_a$. The following relation between the second derivative of the Nekrasov partition function
and the generating function for the $T_{n,n+1}$ knot superpolynomials, holds:
\begin{multline}
\left.\frac{e^{\beta M}}{(1+a)\beta^2}\frac{d^2 Z_{nek}(q,t, m,M,m_a,Q)}{dM\, dm}\right|_{m\to 0,
M\to \infty} \medskip \\ = \sum_n Q^n(tq)^{n/2} P_{n,n+1}(q,t, a)
\end{multline}
This is equivalent to the evaluation of the correlator of local and nonlocal operators defined
above due to the relation
\beq
\left.\frac{e^{\beta M}}{\beta} \frac{\partial Z_{nek}}{\partial M}\right|_{M\to \infty} = \la
\left({\rm Tr} e^{\Phi}\right) \ra,
\eeq
We keep the parameter $a$ in the superpolynomial arbitrary, hence the complete list of the
identifications reads as follows
\beq
a=- e^{-m_a\beta},\quad t=e^{-\beta \epsilon_1}, \quad q=e^{-\beta \epsilon_2}
\eeq
Thus, as above, we have the generating function depending on four fugacities.

The relation between the derivative of the Nekrasov partition function and torus knots
superpolynomials is important per se, however in this paper we are focusing at the critical
behavior in the topological ensembles. Therefore, what we need is i) the identification of the
parameters of the gauge theory as the fugacities of the random walk, and ii) expression of the
generating function for the $q$-Narayana numbers. To make use the known expression for
$q$-Narayana's, we have to switch off one equivariant parameter, $q=t^{-1}$. The critical behavior
is now the $(Q,m_a,q)$--phase space, and in this formulation it is clear that we are dealing with
the 3rd order phase transition. However, the new point is that the 3rd derivative of the Nekrasov
partition function is taken with respect to three distinct variables $(m,M,Q)$ (or, equivalently,
to $(m,M,m_a)$). The physical interpretation of the critical behavior is given in the next Section.

\section{On the physical interpretation of critical behavior}

In this Section we discuss the physical interpretation of the critical behavior found above. As we
have already argued, the critical behavior is exact due to the explicit expression of the
generating function for $q$-Narayana's, which depends on three fugacities (chemical potentials)
controlling length of the path, area under the path and corners. The interpretations of the
fugacities and of the critical behavior are different at Brownian excursion side, knot side, and 5D
SQED side, so we consider them separately. However, these seemingly different critical behaviors
reflect one and the same generic pattern of the phase transition.

\subsection{Critical behavior at the Brownian excursion side}

\begin{figure}
\centerline{\includegraphics[width=8cm]{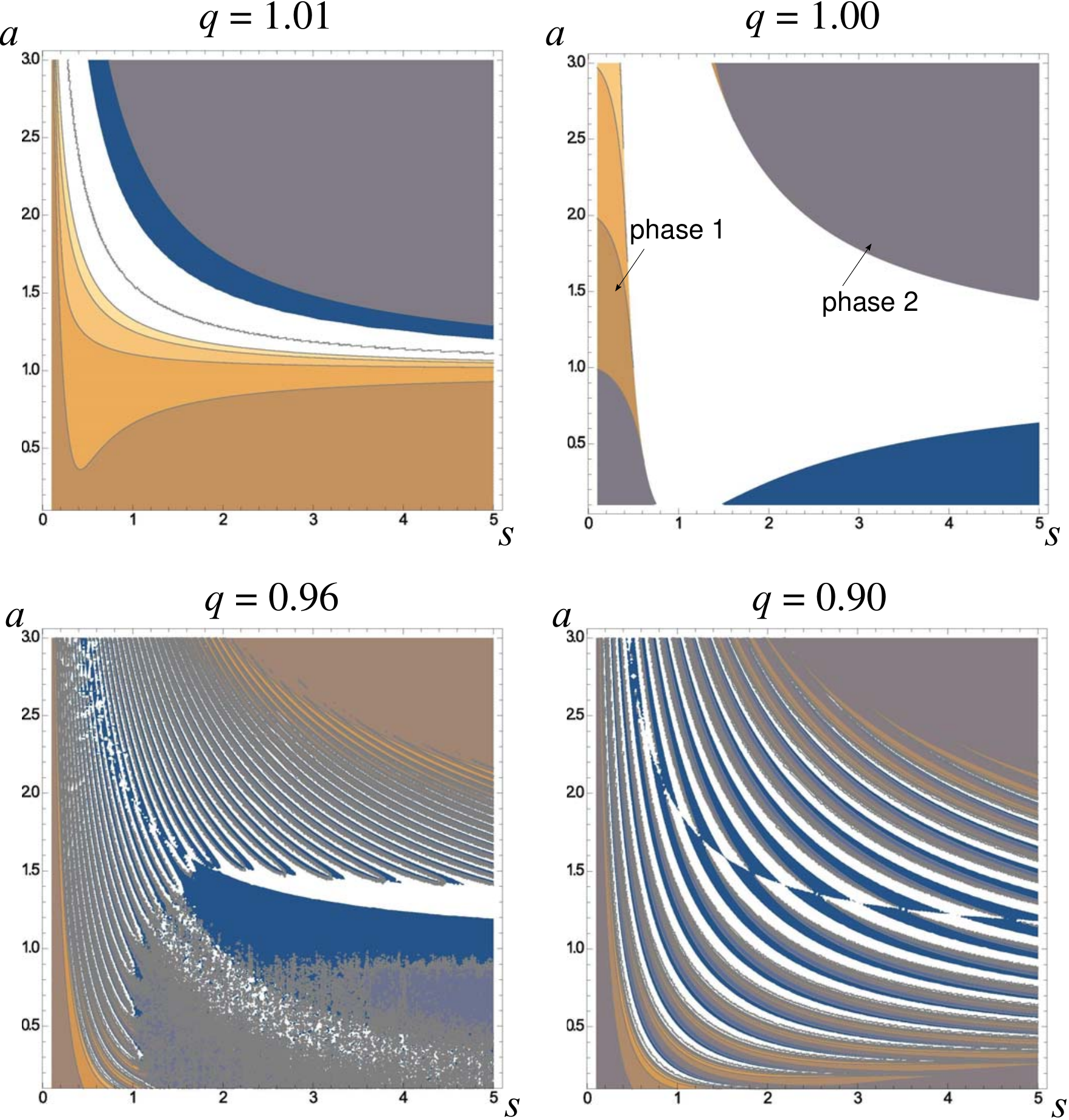}}
\caption{Relief of the generating function $F(a,s)$ above the $(s,a)$--plane for different fixed
values of the area fugacity, $q$. For $q=1$ the phase 1 corresponds to short paths, and the phase 2
-- to long and rather wrinkled paths. For $q \neq 1$ one sees in the $(s,a)$--plane the cascade of
transitions with the Airy-type asymptotics.}
\label{fig:05}
\end{figure}

In terms of the (1+1)D random walks the critical behavior in the $(s,a,q)$ space has the following
interpretation. In the limit $q\rightarrow 1$ the fugacity of the area is switched off and we deal
with the generating function $F(a,s)$ (see \eq{eq:nar}) defined in the $(s,a)$--plane. At large
finite lengths, $n$, the average number of corners, $\la C(n)\ra$ diverges with the length, $n$, as
$\la C(n)\ra\big|_{n\gg 1} = \frac{n}{2}$. The corresponding contour plot of the generating
function $F(s,a)$ is shown in the \fig{fig:05} in the $q=1$--panel. The critical lines divide the
$(s,a)$--plane in three domains. In the first domain ("phase 1") since $s$ is small, one has short
and (in average) sufficiently wrinkled paths, with varying number of corners controlled by the
fugacity $a$. In the second (intermediate) domain the trajectories are long since they exceed the
critical value for the fugacity, again with varying number of corners, however since the generating
function $F(s,s)$ diverges, currently we have not any physical interpretation of this phase. In the
third domain ("phase 2") the trajectories are long and essentially wrinkled, since for any $s$ the
value of $a$ is bounded from below: $a>a_{\rm min}=\frac{1+\sqrt{s}}{s}$.

For $q \neq 1$ the behavior of the system becomes much more rich and the ensemble of area- and
corner-weighted Dyck paths exhibits the cascade of Airy-type (3rd order) phase transitions as it is
seen from the contour plots for the function $F(q,s,a)$ drawn at few fixed values of $q$ ($q=0.9,
0.96, 1.01$) -- see the corresponding panels in the \fig{fig:05}.

\begin{figure}
\centerline{\includegraphics[width=8cm]{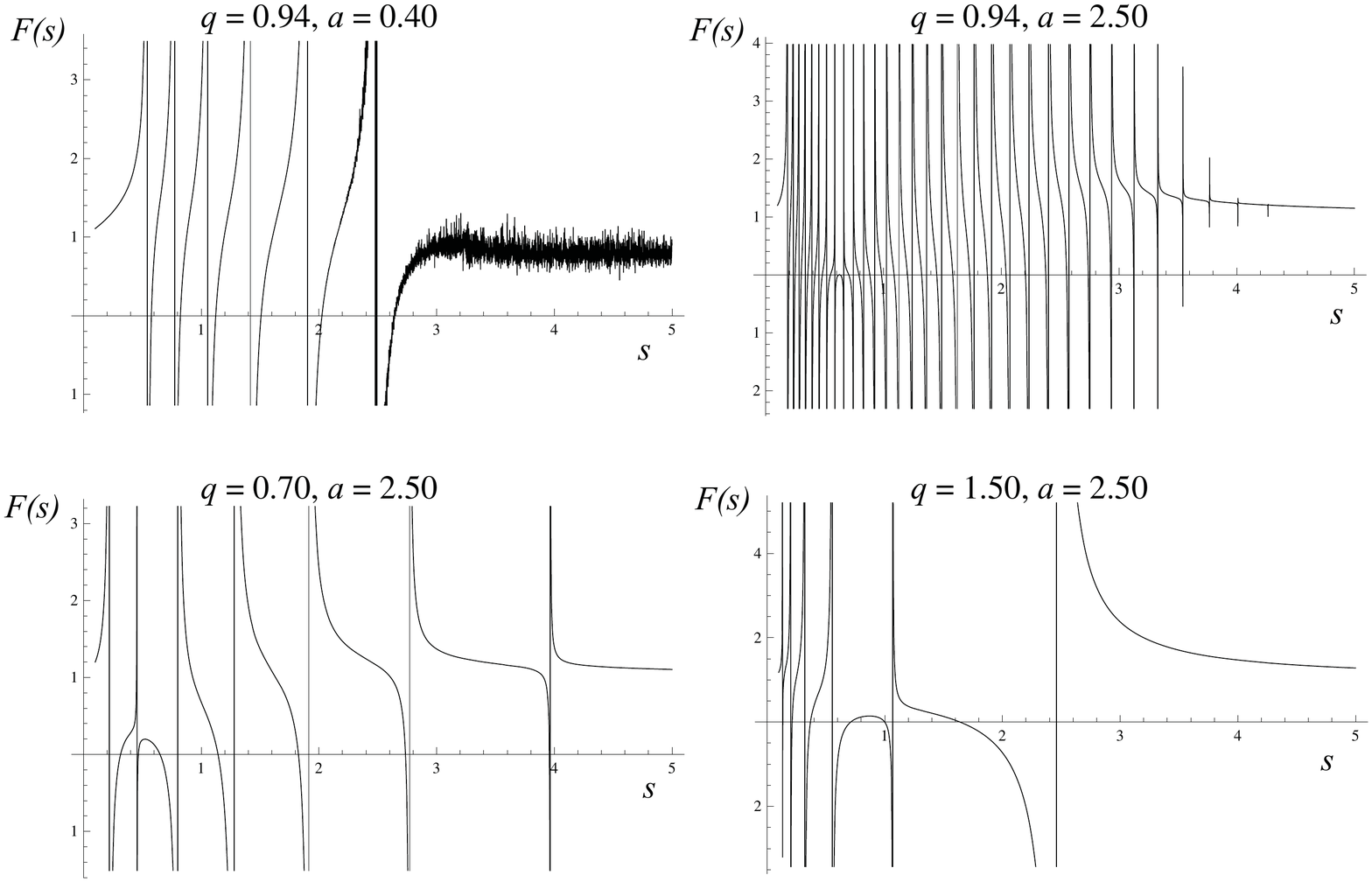}}
\caption{Plots of the generating function $F(s)$ for few fixed values of $q$ and $a$.}
\label{fig:06}
\end{figure}

The asymptotic behavior of the function $F(q,a,s)$ as a function of $s$ for fixed values $q$ and
$a$ near the singularities (transition points) is better seen in the figure \fig{fig:06}. One
clearly distinguishes the finite cascade of phase transitions at different $q\neq 1$ and $a$.

\begin{figure}
\centerline{\includegraphics[width=8cm]{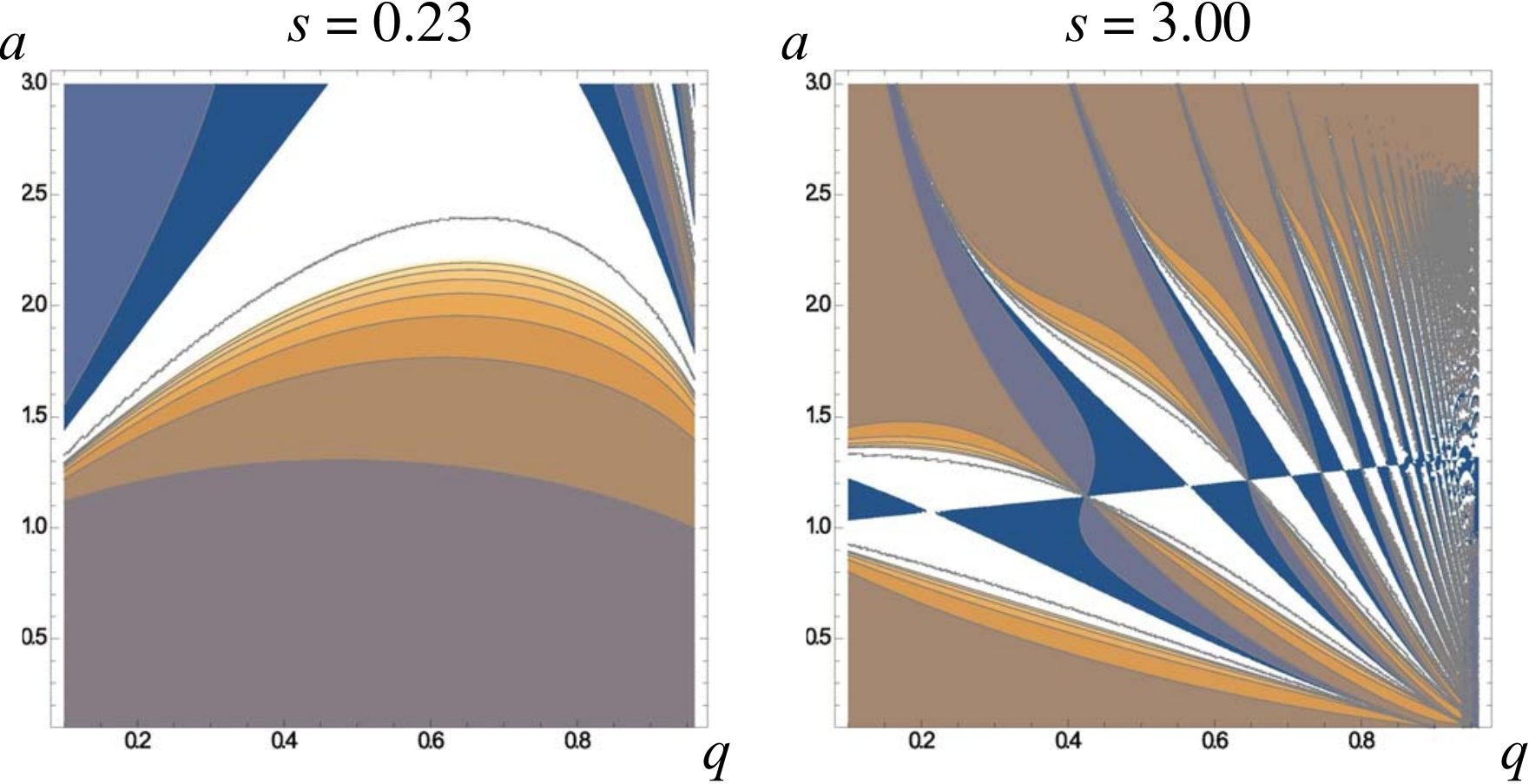}}
\caption{Plots of the generating function $F(q,s)$ above the $(q,s)$ plane for fixed values
$a=0.23, 3.00$. The cascade of transitions in clearly seen.}
\label{fig:07}
\end{figure}

For completeness, we provide in the \fig{fig:07} two contour plots of the function $F(q,s)$ at two
fixed values of corner fugacity, $a=0.23$ and $a=3.00$, which correspond to weakly and highly
wrinkled paths.

The nature of existence of the cascade of transitions lies in the lattice origin of the problem and
deals with the incommensurability of "optimal" trajectories with respect to the lattice spacing for
particular values of $a$ and $q$ and $s$. Actually, it is very easy to imagine that there are
conflicts between wrinkled (with large number of corners) and "inflated" (with extended area)
trajectories at some specific length. On the lattice this competition leads to a finite number of
phase transitions, manifested in the divergence of the generating function $F(q,a,s)$.

\subsection{Critical behavior at knot side}

In the ensemble of the $T_{n,n+1}$ uncolored torus knots, the situation when we can use the exact
answer for the $q$--Narayana's corresponds to the unrefined case with 3D space of parameters
$(s,a,q)$. Hence, we consider generating function for HOMFLY polynomials, $H_{n}(a,q)$, for $T_{n,
n+1}$ torus knots.

What could be the interpretation of the phase transitions in the ensembles of the torus knots?
Naively, different behavior of the generating function at different ridges of the transition point,
means different ability to recognize a particular knot in the ensemble. Speaking more formally, we
could try to distinguish phases via the derivatives of the generating function with respect to the
particular generating parameters, that is the average values of the corresponding "topological
numbers". The derivative with respect to the variables yields the average number of winding, or in
other terms, the type of the torus knot. However the interpretation of other topological numbers is
less evident. On the other hand, these quantum numbers are quite transparent if we remember the
that knot invariants are related to the enumeration of BPS states. Along this approach, other
generating parameters count the number of the corresponding branes.

Let us remind, that in a pure "knot framework", one can obtain the $SU(n)$ invariants of the knots
upon the substitution
\beq
a=q^n
\eeq
Hence, the $q \to 1$ limit corresponds to the situation when all $SU(n)$ invariants are
"degenerate" and cannot be distinguished. At the critical line in the $(a,s)$ plane in the ensemble
of torus knots, the average winding $\la n \ra$ diverges and the average value $\la \frac{dH}{da}
\ra$ diverges as well. On the critical line they are related as
\beq
\la n \ra\propto \la \frac{dH}{da} \ra
\eeq
Hence, in the ensemble of torus knots, the stable limit ($n\to\infty$) is reached at the critical
line, where we also forget about the "nonabelian character" of knot invariants.

Note that the counting problem can be formulated in terms of the Seifert surface, which for the
$T_{n,n+1}$ torus knots reads as
\beq
x^n = y^{n+1}
\eeq
where $x,y \in C$. For instance, the Alexander polynomials can be formulated in terms of the
monodromy in the cohomologies related to the Seifert surface. Also, the Alexander polynomials can
be interpreted in terms of the stratification of the algebra of functions on the Seifert surface.
Hence, the critical behavior can be formulated in terms of the Seifert surface as well. Whether one
can relate the type of the Seifert surface with the singularity type near the transition point of
the generating function, is still an open question.

Let us comment on the interpretation of the cascades of the phase transitions at "torus knot side".
Assuming that there are infinite number of phases in the generic situation, it is naturally to
conjecture the domination of the different values or windings in different phases.

\subsection{Critical behavior at instanton side}

Turn finally to the third part of our construction. In the ensemble of instantons, the account of
three fugacities $(s,a,q)$ corresponds to the account of the coupling constant, mass and parameter
of the $\Omega$--deformation in the unrefined limit. The critical cascades in this case can be
derived again from the analytic structure of the $q$--Narayana generating function, $F(q,s,a)$.

The limit $q\to 1$ corresponds to switching off the $\Omega$--deformation. Hence, the
$(s,a)$--phase space corresponds to the space of the mass, $m_a$, of the hypermultiplet in the
antifundamental representation \cite{gm}, and the instanton charge counting parameter, $Q$. The
critical line in the physical $(s,a)$--variables is defined by the equation
\beq
\left(1+Q-Qe^{m_a\beta}\right)^2=4Q
\label{eq:Q}
\eeq
There are two critical lines, $Q=\left(\frac{1-e^{m_a/2}}{1-e^{m_a}}\right)^2$ and $Q=\left(\frac{1
+ e^{m_a/2}}{1-e^{m_a}}\right)^2$, defined by \eq{eq:Q} at the particular relation between the
gauge coupling and mass of the antifundamental representation. Configurations with different number
of instantons dominate in the different regions of the $(s,a)$--space. At the critical line we have
the following scaling behavior
\beq
\frac{d^2 Z}{dM\,dm_a} \propto \frac{d^2Z}{dM\,dQ}
\eeq

When the parameter of the $\Omega$--deformation is switched back again, the cascade of the singular
surfaces emerges, dividing the three-dimensional parameter space in the regions. Since the
parameter $q$ corresponds in the 5D SQED to the fugacity for the angular momentum in some plane in
$C^2$, the critical surfaces separate the regions with the different responses of our observable to
the rotation. When $a=0$, the antifundamental becomes massless and we have the critical cascades in
$(s,q)$--plane; their scaling is dictated by the analyitic structure of the $q$-Airy functions.

In the simplest case when only the $s$--parameter matters, the critical behavior of the
$q$--Catalans corresponds to the following values of parameters in the 5D gauge theory
\beq
\epsilon_1 = 0,\quad \epsilon_2\beta \to 0,\quad  \ln s = -8\pi^2 \beta g^{-2} = 4\pi \log 2
\eeq

The general issues concerning the 3rd order phase transitions will discussed in the separate
publication \cite{bgmn}.

\section{Area-weighted Dyck paths and Toda system}

Let us point out that the scaling function $C(z)$, which appeared many times throughout the text
(see, for instance, \eq{eq:asymp1}, or \eq{ricatti2}), plays also a very important role, connecting
BE to the integrable systems. It is known (see \cite{fla,rich3,janson} and the references therein)
that $w(z)$, defined in \eq{ricatti2}, is itself a generating function:
\beq
w(z)\Big|_{z\to\infty} \sim \sum_{k=0}^{\infty}\frac{(-1)^k}{2^k\,k!}\, \Omega_k\, z^{-(3k-1)/2},
\label{eq:gen2}
\eeq
where the coefficients $\Omega_k$ have well-defined physical sense: representing $\Omega_k$ in the
form
$$
\Omega_k=2^{(3k-1)/2} \Gamma((3k-1)/2) \la{\cal B}^k\ra
$$
where $\Gamma(...)$ is the gamma-function. One can show \cite{majumdar,janson} that $\la{\cal
B}^n\ra$ is the $n$th moment of the area under the Brownian excursion on the unit interval.
Defining $K_n=\frac{\Omega_n} {2^{n+1}n!}$ one sees that $K_n$ satisfy the recursion
\beq
K_n=\frac{3n-4}{4}K_{n-1}+\sum_{k=1}^{n-1}K_k\, K_{n-k}; \quad K_0=-\frac{1}{2}.
\label{eq:rec3}
\eeq
On the other hand, the function $w(x)$ appears it the theory of integrable systems as the solution
of the rational Painlev\'e II equation ($w(x)\equiv u(x)$) at $\alpha=0$:
\beq
u''(x)=2u^3(x)+ 4x u(x)+ 4\left(\alpha+\frac{1}{2}\right).
\label{eq:painleve}
\eeq

The connection of area-weighted generating function $w(x)$ with the rational solutions of
Painlev\'e II is not restricted exclusively by \eq{eq:painleve}, and can be pushed for any
$\alpha=N+\frac{1}{2}$, where $N=0,1,2,...$. Take into account that the rational solutions of
\eq{eq:painleve} can be written (see \cite{kajiwara1,kajiwara2}) as $u(x)=-\ln\frac{\sigma_{N+1}}
{\sigma_N}$, where $\sigma_N\equiv \sigma_N(x)$ is the $\tau$-function of the Toda system (see
\cite{toda-bilinear2}), written as a Hankel determinant
\beq
\sigma_N=\det \left(\begin{array}{cccc}
a_0 & a_1 & \cdots & a_{N-1} \medskip \\
a_1 & a_2 & \cdots & a_{N} \medskip \\
\vdots & \vdots & \ddots & \vdots \medskip \\
a_{N-1} & a_N & \cdots & a_{2N-2} \end{array} \right)
\label{eq:det1}
\eeq
and the entries $a_n\equiv a_n(x)$ satisfy the recursion \cite{kajiwara2}
\beq
a_n = 2(n-2) a_{n-3} +\sum_{k=0}^{n-2} a_k\, a_{n-k-2},
\label{eq:rec2}
\eeq
with $a_0=x,\; a_1=1,\; a_2=x^2$. The associated generating function,
\beq
G(x,s)=\sum_{j=0}^{\infty}a_j(x) (-2s)^{-j},
\label{eq:gen1}
\eeq
obeys the Riccati equation \cite{kajiwara2} (compare to \eq{ricatti})
\beq
-2 G + G^2 + \partial_s G - (4s^2+s^{-1})G + 4 x s^2=0,
\label{ricatti3}
\eeq
whose solution is
\beq
G(x,s) = 2s^2 + \frac{d}{ds} \ln {\rm Ai}(s^2-x).
\label{eq:airy2}
\eeq

The equation \eq{eq:rec2} at large $n$ resembles (though being different in details) the recursion
\eq{eq:rec3} for the function $K_n$. The connection between \eq{eq:rec3} and \eq{eq:rec2} can be
set by comparing \eq{eq:gen2} and \eq{eq:gen1}. Finally, we get
\beq
a_j=(-2)^j\sum_{k=0}^{\infty}\frac{(-1)^k\Omega_k}{2^{k-1}\,k!} \sum_{j=0}^{\infty}
\left(\begin{array}{c} \frac{1-3k}{2} \\ m \end{array} \right) (-x)^{m} \delta_{3k+2m+1,j},
\eeq
where $\delta_{i,j}$ is the Kronecker $\delta$--function. Thus, we explicitly see how the linear
combinations of moments of area-weighted Brownian excursions, $\Omega_k$, are connected to the
coefficients $a_j$ in the expansion of the Toda $\tau$--function.

Let us also point out the striking similarity of the recursion equation \eq{eq:rec3} for different
momenta of area-weighted Dyck paths with the summation over genus, $g$, the partition function of
the $U(n)_k\times U(n)_{-k}$ Chern--Simons--matter theory, also known as the ABJM theory
\cite{ABJM}. We plan to discuss this question in details in the forthcoming paper \cite{gmn}.

\section{Conclusion}

In this paper, the consideration of the directed random walks with the "area+length+corners"
statistics allows us to find a new critical behavior in the ensemble of the $T_{n,n+1}$ torus knots
and in the ensemble of instantons in 5D SQED. This critical behavior takes place in the space of
fugacities and the corresponding parameters in the gauge theory. The result is exact due to the
explicit answer of the generating function for paths with these statistics.

In the space of fugacities there are patterns of critical lines, and cascades of the phase
transitions. The physics of the phase transitions at this critical line for the ensembles of torus
knots and instantons is very rich. In both cases at the critical line the contribution to the
quantity under consideration with the infinitely large topological number starts to dominate. These
quantities are windings in the knot case, or topological charges in the 5D SQED. In the gauge
theory we have the critical lines in the (gauge coupling, mass) parameter plane. The phase
transition in this plane is physically meaningful and certainly deserved detailed investigation.

Even more profound behavior is seen when all three fugacities are nontrivial. The critical cascades
are quite unexpected and physically they take place in the unrefined limit of the
$\Omega$--deformed gauge theory. At the knot side, the critical behavior occurs in the generating
function of the $T_{n,n+1}$ knots characterized by the HOMFLY polynomials.

The following general comment is in order. One has to distinguish two "critical phenomena": phase
transition and the wall--crossing phenomena. The latter is typical for the supersymmetric gauge
theories when the spectrum of BPS states jumps at some curves and surfaces in the moduli space.
Upon the SUSY breaking, the curves of marginal stability become the curves of phase transitions.
Hence literally the wall-crossing should not be considered as a phase transition, since only the
spectrum of the stable BPS states gets rearranged. However, there is a pattern of real phase
transitions, even in SUSY theory, with the Argyres--Douglas superconformal point in SQCD as the
notable example. In this case the particular condensates serve as the order parameter.

Do we deal with the phase transition or with the wall-crossing phenomena in our case? Formally, the
3rd derivative of the Nekrasov partition function in 5D SQED diverges, hence we could speak about
the 3rd order phase transition in the instanton ensemble. Moreover, the UV properties of the
condensate play the key role, and critical lines involve the hypermultiplet masses and the gauge
coupling. Hence, we could speak on some similarity with the Argyres--Douglas scenario. On the other
hand, at the torus knot side, we could remind the interpretation of the knot superpolynomial as the
index of the BPS states in some theory. Hence the jump in the index is better interpreted in terms
of wall-crossing.  As far as we know, there is no evident notion of the wall-crossing in terms of
random walks, however, we cold think about it in terms of the Stokes lines in quantum mechanics.
Summarizing, we think that the critical behavior we have described, admits the interpretation both
as the wall-crossing, and as the phase transition, depending on the viewpoint.

The critical behavior we have considered in 5D SQED is a nontrivial example of the 3rd order phase
transition. The familiar example of the 3rd order phase transition in the gauge theories is the
phase transition of Douglas--Kazakov type \cite{kazakov} in 2D pure Yang-Mills theory on the sphere
characterized by the singularity \eq{eq:asymp1}. Somewhat similar phase transition of
Douglas--Kazakov type has been observed in \cite{nekmar}. In all these situations the phase
transitions are driven by instantons. On the other hand, our study, as well as \cite{gm}, yield the
clear relation between the torus knots and instantons. This suggests more general viewpoint that
there is a knotting process behind the 3rd order phase transitions in the general case and the knot
invaraints play the role similar to the central charges of the Virasoro algebra in the 2nd order
phase transitions. We shall provide the general discussion of the 3rd order phase transition in the
gauge theories, knot ensembles and statistical mechanics elsewhere \cite{bgmn}.

There are many challenging issues, related to our work, which deserve further study. For example,
it would be interesting to extend the analysis to the whole ensemble of $T_{n,m}$ torus knots and
the whole set of fugacities. Also, it would be important to recognize the counterparts of the
"bounces" and "corners" in the particle path integral in the continuum (i.e. off lattice) and
obtain the interpretation of the external magnetic field in the BE approach as a kind of the Berry
curvature from a 4D viewpoint. It would be also very interesting to extend our consideration and to
relate the spectrum of the full (i.e. non-chiral) Hofstadter model at BE side with knot invariants
and 4D instantons. We plan to address all these questions in \cite{bgmn}. It seems also very
important to recognize the critical behavior observed in this paper in the ensembles of branes
\cite{bg}, Hopfions \cite{nitta} and "ensemble" of 3D theories classified by the torus knots
\cite{gukov2014}. Another interesting question concerns the physical interpretation of the critical
behavior in terms of the extended objects involved in the torus knot description along the
discussion in \cite{gm,gm2}. This is the explicit realization of the $S$-dual magnetic approach to
the evaluation of the knot invariants suggested in \cite{witten}.

We are grateful to E. Gorsky, A. Milekhin  and N. Nekrasov for the useful discussions. The work of
A.G. and K.B. was supported in part by grants RFBR-12-02-00284 and PICS-12-02-91052. The work of
K.B. was also supported by the Dynasty fellowship program and Princeton Centennial Fellowship. A.G.
thanks SCGP at Stony Brook University where the part of this work has been done during the Simons
Workshop on Mathematics and Physics 2014 for the hospitality and support.

\begin{appendix}

\section{$q$-statistics of Dyck paths in a group-theoretic setting}

\subsection{Area-weighted Dyck paths as random walks on a $q$-deformed locally-free semigroup}
\label{sect:2:2}

\begin{figure}[ht]
\centerline{\includegraphics[width=6cm]{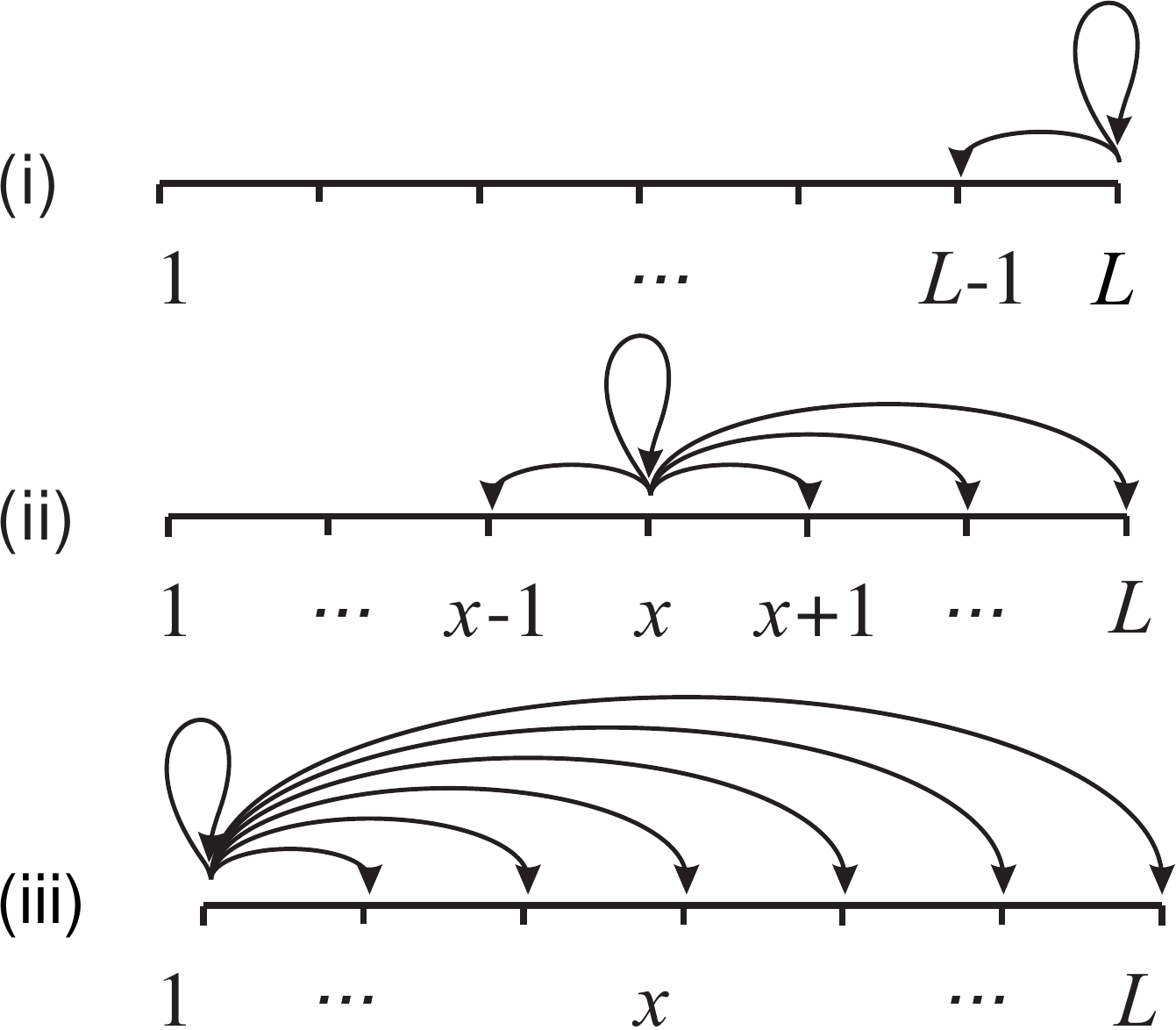}}
\caption{(a) The set of possible values which can take the index $y$ if the previous index is $x=0$
(i), $x=1,2,...L-1$ (ii) and $x=L$ (iii).}
\label{fig:08}
\end{figure}

Let us associate each entrance of a Dyck path at a height $x$ with the application of a "generator"
$\hat{g}_x$. Define the semigroup $F^+$ with the infinite set of generators $\{\hat{g}_0,
\hat{g}_1, \hat{g}_2,...\}$, obeying the commutation relations
\beq
\hat{g}_j \hat{g}_k = \hat{g}_k \hat{g}_j \quad \forall \; |k-j|\ge 2
\label{a:08}
\eeq
Any Dyck path can be uniquely encoded by a "word" written in terms of generators $\hat{g}_x$
($x=1,2,...$). For example, the Dyck path in the \fig{fig:03} corresponds to a word
\beq
{\cal W}=\hat{g}_0\,\hat{g}_3\,\hat{g}_2\,\hat{g}_1\,\hat{g}_2\,\hat{g}_1\,\hat{g}_3\,\hat{g}_3\,
\hat{g}_2\, \hat{g}_1\, \hat{g}_1\,\hat{g}_0\, \hat{g}_3\,\hat{g}_2\,
\hat{g}_1\,\hat{g}_1\,\hat{g}_1\,\hat{g}_0
\label{a:09}
\eeq
(see also the \fig{fig:04}). Define a "normal order" representation of words in the semigroup
$F^+$, which consists in pushing the generators with smaller indices \emph{as left as possible} if
such a reordering does not violate the commutation relations \cite{vershik}. Let us note that all
Dyck paths written in terms of generators $\hat{h}_x$ ($x=1,2,...$) are automatically normally
ordered. Thus, we can compute the partition function of all such Dyck paths via the transfer matrix
approach, where the transfer matrix, $R$, defines which particular generator, $\hat{g}_y$, can stay
next to the previous one, $\hat{g}_x$:
\begin{itemize}
\item[(i)] If $x=0$ then $y=0,1,...,n$;
\item[(ii)] If $x\ge 1$ then $y=x-1, x, x+1,...,n$;
\item[(iii)] If $x=n$ then $y=n$
\end{itemize}
Graphical representation of the rules (i)-(ii) is shown in the \fig{fig:08}.

In order to count heights, identify a weight $q^x$ with the generator $\hat{g}_x$. Then the
weighted transfer matrix, $R(q)$, consistent with the rules (i)-(iii) has the following form
\beq
R(q)=\left(\begin{array}{ccccc}
1 & q & q^2 & q^3 & \ldots \\
1 & q & q^2 & q^3 & \ldots \\
0 & q & q^2 & q^3 &  \\
0 & 0 & q^2 & q^3 &  \\
\vdots & \vdots &  &  & \ddots \end{array}\right)
\label{a:10}
\eeq
Define ${\bf W}_n(q)=\big(W_n(0;q),W_n(1;q),W_n(2;q),...\big)^{\intercal}$. The number of Dyck
paths, starting at the height $x=0$ and returning to the height $x=0$ is given by the element
$W_n(0;q)$ of the vector ${\bf W}_n(q)$, where
\beq
W_n(0;q) = {\bf V}\, R^n(q)\, {\bf V}^{\intercal}; \qquad {\bf V}=(1,0,0,...)
\label{a:11}
\eeq
One can straightforwardly check that
\beq
W_n(0;q) = C_n(q)=Z_{2n}(1;q)
\label{a:12}
\eeq
where $C_n(q)$ is the $n^{\rm th}$ Carlitz--Riordan $q$-Catalan number and $Z_n(1;q)$ is given by
\eq{eq:03}. Thus, the magnetic random walks can be equivalently described either by the transfer
matrix $T(q)$ (Eq.\eq{eq:02}), or by the transfer matrix $R(q)$ (Eq.\eq{a:10}). The last
description can be straightforwardly generalized to account for corners.

\subsection{Area- and corner-weighted Dyck paths and $(q,a)$-deformed locally free semigroup}

The approach developed in the Section \ref{sect:2:2} can be easily extended on counting of
simultaneously area-weighted and $\wedge$-corner-weighted $2n$-step Dyck paths shown in the
\fig{fig:04}. Specifically, we are interested in calculation of the partition function, ${\cal
Z}_{2n}(q,a)$, defined in \eq{eq:1}, where the summation runs over the ensemble of $2n$-step Dyck
paths enclosing the area $A$, controlled by the fugacity $q$, and having $M$ $\wedge$-corners,
controlled by the fugacity $a$ (bounces are not counted, so we set $t=1$ in \eq{eq:1}).

Note that the $\wedge$-corner appears in position $k$ if and only if the following condition is
fulfilled: $h_{k+1}\leq h_k$. In the opposite situation, i.e. for $h_{k+1}=h_k-1$, the
$\wedge$-corner is not created. In terms of the transfer matrix approach, which counts the normally
ordered words in the semigroup $F^+$, the $\wedge$-corner creation condition means that for all
steps which do not decrease the height we should apply the weight $a$ (the fugacity of the corner).
Thus, we get the following $(q,a)$-deformed transfer matrix $R(q,a)$, where we have multiplied all
entries of matrix $R(q)$, except the lower sub-diagonal, by a factor $a$:
\beq
R(q,r)=\left(\begin{array}{ccccc}
a & aq & aq^2 & aq^3 & \ldots \\
1 & aq & aq^2 & aq^3 & \ldots \\
0 & q & aq^2 & aq^3 &  \\
0 & 0 & q^2 & aq^3 &  \\
\vdots & \vdots &  &  & \ddots \end{array}\right)
\label{a:14}
\eeq
(compare to \eq{a:10}). Evaluation of the area- and corner-weighted partition function, ${\cal
N}_n(q,a)$ is similar to \eq{a:11}, namely:
\beq
{\cal N}_n(q,a)\equiv Z_{2n}(q,a)={\bf V}\, R^n(q,a)\, {\bf V}^{\intercal};\;  {\bf V}=(1,0,0,...)
\label{a:15}
\eeq

In particular, for ${\cal N}_n(q=1,r)$ we get the Narayana polynomial, which is the generating
function for the Narayana numbers ${\cal N}_{n,k}$,
\beq
{\cal N}_n(q=1,a) = \sum_{k=1}^{k} {\cal N}_{n,k}a^k = ~_2F_1(1-n,-n,2,a)\,a;
\eeq
where $_2F_1(1-n,-n,2,a)$ is the hypergeometric function and
\beq
{\cal N}_{n,k}=\frac{1}{n}\left(\begin{array}{c} n \\ k\end{array}\right) \left(\begin{array}{c} n
\\ k-1 \end{array}\right)
\label{a:16}
\eeq
For example, for $n=7$ both Eq.\eq{a:15} (in which we take $q=1$) and Eq.\eq{a:16} give
\beq
Z_{2n}(q=1,a) = a + 21 a^2 + 105 a^3 + 175 a^4 + 105 a^5 + 21 a^6 + a^7
\eeq
meaning that in the ensemble of $n=14$--step Dyck paths has 21 different configurations with 2
$\wedge$-corners, 105 configurations with 3 $\wedge$-corners etc.

For any $(q,a)$ the recursion relation for ${\cal N}_n(q,a)$ reads
\beq
{\cal N}_n(q,a) = a {\cal N}_{n-1}(q,a)+\sum_{k=1}^{n-1}q^k {\cal N}_k(q,a){\cal N}_{n-k-1}(q,a)
\label{a:17}
\eeq
(compare to \eq{eq:04}). The generating function
$$
F(q,a,s)=\sum_{n=0}^{\infty} s^n {\cal N}_n(q,a)
$$
obeys the functional relation
\beq
F(q,a,s)=1+(a-1)s F(q,a,s)+s F(q,a,s)F(sq,a,s)
\label{a:18}
\eeq
(compare to \eq{eq:05}). These polynomials coincide with the standard $q$-Narayana polynomials
\cite{cigler}. For $q=1$ we reproduce the generating function for Narayana numbers \cite{deutsch}:
\beq
F(a,s)= \frac{1 -(1-a)s - \sqrt{(1-s+sa)^2-4s}}{2s}
\label{a:18a}
\eeq

\subsection{$q$-orthogonal polynomials for area- and corner-weighted Dyck paths}

Consider the recursion relation
\beq
{\bf W}_{j+1}(q,a) = R(q,a) {\bf W}_j(q,a)
\label{a:19}
\eeq
for the vector ${\bf W}_j(q,a)= (W_j(1;q,a),W_j(2;q,a),...)^{\intercal}$, where ${\bf
W}_0(q,a)=(1,1,...)^{\intercal}$ and the transfer-matrix $R(q,a)$ is defined in \eq{a:14}. One can
easily rewrite \eq{a:19} for components of the vector ${\bf W}_j$:
\begin{multline}
W_{j+1}(x;q,a) = q^{x-1}W_j(x-1;q,a) \medskip \\
+ a\sum_{y=1}^L q^{y-1} W_j(y;q,a); \quad 1\le x\le L
\label{a:20}
\end{multline}
For the generating function,
$$
F(x;q,a,s)\equiv \bar{F}(x)=\sum\limits_{j=0}^{\infty} W_j(x;q,a) s^j
$$
we obtain from \eq{a:20}
\beq
s^{-1} \bar{F}(x) - s^{-1} = q^{x-2} \bar{F}(x-1) + a\sum_{y=1}^L q^{y-1} \bar{F}(y)
\label{a:21}
\eeq
which can be rewritten in the local form
\beq
\bar{F}(x+1)=(1+sq^{x-1}(1-a))\bar{F}(x)-sq^{x-2}\bar{F}(x-1)
\label{a:23}
\eeq
This recursion relation looks pretty much similar to the recursion relation for orthogonal
polynomials associated with Rogers--Ramanujan continued fraction -- see \cite{salam}:
\beq
\begin{array}{l}
U_{m+1}(z;b,c) = z(1 + b q^m) U_m(z;b,c) \medskip \\ \hspace{3cm}
- c q^{m-1} U_{m-1}(z;b,c);  \quad (m > 0) \medskip \\
U_0(z;b,c)=1; \quad U_1(z;b,c)=z(1+b)
\end{array}
\label{a:24}
\eeq
The function $U_m(z;c,d)$ has the following explicit expression
\beq
U_m(z;b,c) = \sum_{k=0}^{[m/2]}\frac{(-b;q)_{m-k}\, (q;q)_{m-k}\, z^{m-2k}(-c)^k} {(-b;q)_k\,
(q;q)_k\, (q;q)_{m-2k}}\, q^{k(k-1)}
\label{a:25}
\eeq
where $(b;q)_k=\prod\limits_{j=0}^{k-1} (1-b q^j)$ is the Pochhammer symbol. By setting
\beq
z=1; \quad b=\frac{s(1-a)}{q}; \quad c=\frac{s}{q}; \quad m=x
\label{a:26}
\eeq
we get the following expression for the generating function $F(x;q,a,s)$
\beq
F(x;q,a,s)=U_x\left(1;\frac{s(1-a)}{q}, \frac{s}{q} \right)
\label{a:27}
\eeq
The function $F(x;q,a,s)$ can be represented as a continued fraction of Rogers--Ramanujan
\cite{salam}:
\begin{multline}
F(q,a,s)=\lim_{x\to\infty} F(x;q,a,s)  \\ = \frac{1+\frac{(1-a)s}{q}}{\disp 1+s(1-a)-\frac{s}{\disp
1+s(1-a)q-\frac{s q}{1+s(1-a)q^2 - \frac{s q^2}{1-...}}}} \medskip \\ \hspace{-3cm} \hspace{-3cm}
=\frac{A_q(s;s(1-a))}{A_q(s/q;s(1-a)/q)}
\label{a:28}
\end{multline}
where $A_q(s,s(1-a))$ is the extension of the $q$-Airy function $A_q(s)$ defined in \eq{eq:07}. The
function $A_q(s;s(1-a))$ reads
\beq
A_q(s;s(1-a)) = \sum_{k=0}^{\infty} \frac{q^{k^2}(-s)^k}{(q;q)_k(-s(1-a);q)_k}
\label{a:29}
\eeq
At $a=1$ Eq.\eq{a:29} coincides with $A_q(s)$ in Eq.\eq{eq:07}.

\section{ On relation with the hydrodynamical equations}

In this Appendix we shall mention that the asymptotics \eq{eq:asymp1} describes the scaling of top
line in a bunch of directed vicious walks. Proceeding as in \cite{spohn-pr-fer}, take the ensemble
of $N$ vicious walkers, define the averaged position of the top line and consider its fluctuations
near the averaged position. In such a description all vicious walkers lying below the top line play
a role of a "mean field", which pushes the top line to some "atypical" equilibrium position, around
which it fluctuates. It is naturally to suppose that the fluctuations of the top line in a
mean-field approximation have the same scaling as the fluctuations of the "inflated" Brownian
excursion with fixed area under the path. One actually can show that, following the line of
reasoning of the work \cite{nowak}. The solution of the inviscid Burgers equation
$$
\partial_t u_0(x,t) + u_0(x,t) \partial_x u_0(x,t) = 0
$$
is $u_0(x,t=N) = \frac{x}{2N} \pm \frac{\sqrt{x^2-4N}}{2N}$ and gives the Wigner semicircle law
centered at the point $\frac{x}{2N}$. One can smear the function $u_0(x,t)$ near the boundary
value, $x=x_c = \pm 2\sqrt{N}$, adding by hands the Gaussian fluctuations, i.e. passing to the
Burgers equation with a weak diffusivity ($0<\nu\ll 1$),
$$
\partial_t u(x,t) + u(x,t) \partial_x u(x,t) = -\nu \partial_{xx} u(x,t)
$$
Seeking for weakly fluctuating solutions of viscous Burgers equation near the top line, ($t=N$), in
the form \cite{nowak}
$$
\left\{
\begin{array}{ll}
\disp x=x_c + \nu^{\alpha} y = 2t^{1/2}+ \nu^{\alpha} y; \medskip \\ \disp u(x,t) = \frac{x_c}{2t}
+ \nu^{\beta} w(s,t)= t^{-1/2}+\nu^{\beta} w(y,t) \end{array} \right.
$$
and substituting the ansatz for $u(x,t)$ into the viscous Burgers equation, one gets the equation
for $w(y,t)$, which for $\alpha=2/3$, $\beta=1/3$ and appropriate boundary conditions is
transformed in the limit $\nu\to 0$ into the dimensionless Ricatti equation \cite{nowak}
\beq
-yt^{-3/2}+ \frac{1}{2} w^2 + \partial_y w=0,
\label{ricatti}
\eeq
having the solution (for $t=N$)
\beq
w(z) = 2\frac{d}{dz}\ln {\rm Ai}(2^{-1/3}z) \quad z=yN^{-1/2}.
\label{ricatti2}
\eeq
In \eq{ricatti2} one can recognize the singular part of the grand partition function of the
"area+length"--weighted Brownian excursion. To make this connection precise, define the partition
function $Z_n(A)$ of $n$-step directed 2d random walk in the upper half-plane of the square lattice
(i.e. the Brownian excursion) with the fixed area, $A$. Thus, one can straightforwardly identify
$Z(s,q)$ with $u(y,N)=N^{-1/2}+\nu^{1/3} w(y,N)$ under the following redefinitions:
\beq
\left\{
\begin{array}{rcl}
1-q & \leftrightarrow & \nu, \medskip \\
\frac{1}{4}-s & \leftrightarrow & 2^{-7/3}yN^{1/6}.
\end{array} \right.
\eeq

\end{appendix}

\end{document}